\documentclass[11pt,tightenlines, eqsecnum,aps, prd,preprintnumbers,amssymb, nofootinbib]{revtex4-2}
\usepackage{tikz-feynman}
\tikzfeynmanset{compat=1.0.0}
\usepackage{graphicx}
\usepackage{amssymb,mathtools}
\usepackage{amsmath}
\usepackage{epstopdf}
\usepackage{hyperref}
\usepackage{color}
\usepackage{mathrsfs}

\def\dm2{{d-2 \over 2}}

\def\dcft{\mathcal{D}^{\cft}}
\def\dflat{\mathcal{D}^{\flat}}
\def\A[#1]{A^{(#1)}}
\def\d[#1]{\Delta_{#1}}
\def\D[#1]{\Delta^{(#1)}}
\def\Dv[#1]{\Delta^{[#1]}}
\def\gflata[#1,#2][#3,#4]{\mathcal{G}^{\flat}_{\d[#1],\ldots,\d[#2]}(x_{#3},\ldots,x_{#4})}
\def\dflata[#1,#2]{\mathcal{D}^{\flat}_{\d[#1],\ldots,\d[#2]}(\{r_i\})}
\def\dcfta[#1,#2]{\mathcal{D}^{cft}_{\d[#1],\ldots,\d[#2]}(\{s_i\})}
\def\gcfta[#1,#2][#3,#4]{\mathcal{G}^{cft}_{\d[#1],\ldots,\d[#2]}(x_{#3},\ldots,x_{#4})}
\def\gcfte[#1,#2,#3][#4,#5]{\mathcal{G}^{cft}_{\d[#1],\ldots,\d[#2];\d[#3]}(x_{#4},\ldots,x_{#5})}

\def\dcft{\mathcal{D}^{cft}}
\def\gcft{\mathcal{G}^{cft}}

\def\sumi[#1]{\sum_{i=1}^{[#1]}}
\def\zbar{\Bar{z}}

\def\iflata[#1,#2][#3,#4]{\mathcal{I}^{\flat}_{\d[#1],\ldots,\d[#2]}(x_{#3},\ldots,x_{#4})}

\newcommand{\p}{P}
\newcommand{\db[1]}{\bar{\Delta}_{#1}}
\newcommand{\dt[1]}{\tilde{\Delta}_{#1}}
\newcommand{\Db[1]}{\bar{\Delta}^{({#1})}}
\newcommand{\Dt[1]}{\tilde{\Delta}^{({#1})}}
\newcommand{\mi[1]}{\int \tilde{d #1} }

\newcommand{\hc}[2]{\mathcal{H}^{cft}_{\d[#1],\ldots,\d[#2]}(\{s_{ij}\}_{(#2)})}
\newcommand{\dc}[3]{\tilde{\mathcal{D}}^{cft}_{\d[#1],\ldots,\d[#2]}(a_{1},\ldots,a_{#3})}

\newcommand{\amp}[2]{\mathcal{M}_{\d[1],\ldots,\d[#1]}\left(\left\{#2\right\}_{(#1)}\right)}      
\newcommand{\am}[1]{\mathcal{M}_{\d[1],\ldots,\d[#1]}\left(\left\{P_{ij}\right\}_{(#1)}\right)} 
\newcommand{\amc}[1]{\mathcal{M}^{cft}_{\d[1],\ldots,\d[#1]}\left(\left\{P_{ij}\right\}_{(#1)}\right)}         
\newcommand{\ame}[3]{\mathcal{M}_{\d[1],\ldots,\d[#1];\d[0];\d[#2],\ldots,\d[#3]}\left(\left\{P_{ij}\right\}_{(#3)}\right)}          
\newcommand{\amec}[3]{\mathcal{M}^{cft}_{\d[1],\ldots,\d[#1];\d[0];\d[#2],\ldots,\d[#3]}\left(\left\{P_{ij}\right\}_{(#3)}\right)}   
\newcommand{\ampe}[4]{\mathcal{M}_{\d[1],\ldots,\d[#1];\d[0];\d[#2],\ldots,\d[#3]}\left(\left\{#4\right\}_{(#3)}\right)}

\newcommand{\mdcft}{\tilde{\mathcal{D}}^{cft}}

\newmuskip\pFqmuskip
\newcommand*\pFq[6][8]{%
	\begingroup 
	\pFqmuskip=#1mu\relax
	\mathchardef\normalcomma=\mathcode`,
	\mathcode`\,=\string"8000
	\begingroup\lccode`\~=`\,
	\lowercase{\endgroup\let~}\pFqcomma
	{}_{#2}F_{#3}{\left[\genfrac..{0pt}{}{#4}{#5};#6\right]}%
	\endgroup
}
\newcommand{\pFqcomma}{{\normalcomma}\mskip\pFqmuskip}

\begin{document}
	\title{Feynman Diagrams from Conformal Integrals}
	\author{Siddharth G. Prabhu}
	\email{siddharth.g.prabhu@gmail.com}
	\affiliation{Department of Theoretical Physics, Tata Institute
		of Fundamental Research, Homi Bhabha Rd, \\Mumbai 400005, India}
	
	\begin{abstract}
			We show that momentum space Feynman diagrams involving internal massless fields can be cast as conformal integrals. This leads to a classification of all Feynman diagrams into conformal families, labelled by conformal integrals. Computing the conformal integral in each family suffices to compute all the Feynman diagrams in the family exactly. Using known exact results for some conformal integrals, we present the solutions to the other Feynman diagrams in the corresponding families. These are answers to either finite or dimensionally regularized Feynman diagrams to all orders in the regularization parameter. 			
	\end{abstract}

		\maketitle
	
\section{Introduction}

Primarily, our knowledge of the interactions of elementary particles is determined by the calculations of probabilities of transitions between states, encoded in the Scattering or S-matrix. Typically, S-matrix elements  or more generally correlation functions, are computed in a perturbative series in the coupling, generated by Feynman diagrams. The diagrams are expressed in momentum space as $L-$dimensional integrals, each integral over one of the momenta running through the $L$ loops of the diagram. Naturally, these integrals are functions of Lorentz invariants, i.e squares of momenta, of the particles whose scattering amplitude is being computed. Being of central importance to the connection of the theory of elementary particles with experiments, the computation of the Feynman integrals associated with these diagrams has been an extremely active and fruitful affair, as for example reviewed in \cite{Smirnov:2012gma, Weinzierl:2022eaz}. 

On the other hand, consider correlators in a conformal field theory \cite{Symanzik:1972wj}, or more generally, conformal partial waves and conformal blocks \cite{Ferrara:1972kab, Ferrara:1974nf, Ferrara:1974ny, Dolan:2000ut, Dolan:2003hv, Dolan:2011dv}. All these conformal objects can be expressed as conformal integrals. Such integrals being conformally covariant can be expressed, upto a prefactor, as functions of conformal cross ratios.

 In this article, we show that every Feynman integral with massless internal fields, and massless or massive external ones, is a limit of such a conformal integral. In \cite{Prabhu:2023oxv}, it was shown  that position space correlators can be cast in a conformal representation that leads to such conformal integrals. Translating this position space analysis to momentum space provides the central results of this article.  Position space correlators are defined as integrals over spacetime locations of vertices of the Feynman diagram. The conformal representation is achieved by adding an extra field at each vertex of the position space Feynman diagram such that the sum of powers of propagators at each vertex is equal to the spacetime dimension. We call this latter condition the conformal constraint. Taking all the extra fields of the conformal diagram to infinity gives the answer to the general Feynman integral one began with.

Here, we use the fact that every loop integral in momentum space is formally similar to its counterpart in position space, with the integral over a loop momentum being the analogue of an integral over the spacetime location of a vertex. The associated conformal diagram is obtained by adding an extra internal line to each loop of the original diagram such that the sum of powers of propagators belonging to each loop is equal to the spacetime dimension. Collapsing all the extra lines in every loop takes us from the conformal diagram result to the general Feynman diagram. This naturally gives rise to an organization of Feynman diagrams into conformal families, each of which has a conformal diagram. Collapsing different internal lines of this conformal diagram gives rise to the other (non-conformal) members of the family.

Scattering amplitudes in planar $\mathcal{N}=4$ SYM are known to enjoy dual conformal invariance \cite{Drummond:2006rz,Bern:2006ew,Drummond:2007aua}, which is seen by going from the momenta variables to dual momenta variables. In the dual variables, one lands at conformal integrals which are particular examples of the ones we encounter here. However, our result connecting conformal integrals to general Feynman integrals holds for all Feynman diagrams, including non-planar ones. 

Let us now define the principal characters in our story.
The momentum space representation of Feynman integrals involving internal massless fields can be written as
\begin{equation} \label{feynint}
	\begin{split}
		\prod_{r=1}^{L} \int \frac{d^d \ell_r}{\pi^{d/2}} \prod_{i=1}^{I}\frac{\Gamma(\d[i])}{  \left\{ \left(\sum_{j=1}^L A_{ij} {\ell}_j + p_i \right)^2  \right\}^{\d[i]}}
	\end{split}
\end{equation}
Such an integral can represent a single Feynman diagram or a sum of Feynman diagrams or an auxiliary integral that appears enroute to the computation of Feynman diagrams. Here, by either Feynman diagram or Feynman integral, we will mean an expression of the kind in \eqref{feynint}. For example, consider massless scalar field theory with a  $\phi^n$ interaction. Every $L-$ loop diagram with momentum $l_r$ running in the $r^{th}$ loop, and $I$ internal lines is such an integral with all $\d[i]$, which we shall call scaling dimensions, equalling unity. The momentum running in each internal line is of the form $A_{ij} {\ell}_j + p_i$ for some constants $A_{ij}$ and momenta $p_i$. When momentum conservation is imposed at each vertex, as is usually done, the momenta $p_i$ get linearly related to the external momenta flowing into the diagram. Here, we shall work in Euclidean space and not impose momentum conservation, with the understanding that momentum conservation is only demanded of the analytically continued Minkowskian counterpart\footnote{This helps with the analytic continuation because momentum conservation is more restrictive in Euclidean as compared to Minkowskian space, for example setting some Euclidean momenta to zero whose Minkowskian counterparts would be non zero.}. As a result, we will draw diagrams without any external fields so that our diagrams actually represent any member of the collection of Feynman diagrams with the same internal structure and varying number of external fields at each vertex. 

The integrals in \eqref{feynint} with general $\d[i]$ appear in various contexts (see, for example \cite{Smirnov:2012gma,Weinzierl:2022eaz}) such as when reducing diagrams associated with tensor fields to scalar integrals. The widely used and powerful integration by parts method \cite{Tkachov:1981wb,Chetyrkin:1981qh} relates Feynman diagrams sharing a common set of internal lines but with different values of $\d[i]$, and hence, also necessitates the computation of the integrals with general scaling dimensions.

In the next section, we find that every $1-$loop diagram is in fact equivalent to a conformal integral. In \S \ref{sec:lloop}, we describe conformal families at $2-$loop and higher loops. Then, we study some particular examples of the general relations we have described. In \S\ref{sec:lad}, we consider ladder diagrams in four dimensions. From the known $L-$ loop four point ladder diagrams, we immediately find their 3-point and 2-point counterparts. In \S\ref{sec:conffam}, we write down general dimensional answers to Feynman diagrams in the first few families, namely those of the conformal box, the conformal pentagon and the conformal double box. In the latter case, and more generally, we lay out the procedure to find on-shell Feynman diagrams, and in particular ones with massive external fields from conformal diagrams. All the analysis in the main text is for Feynman diagrams in momentum space. For actual computations, we use position space. In the appendices, we state some results for position space Feynman diagrams, compile relevant results from \cite{Prabhu:2023oxv}, and also define all the notation used in the main text.

\section{Families of 1-loop diagrams}\label{sec:1l}
Consider the most general one-loop diagram with $n$ internal lines,
\begin{equation} \label{1loop}
{\cal M}_{\d[1],\ldots,\d[n]}(p_1,\ldots,p_n) = \mathcal{A}\int \frac{d^d \ell}{\pi^{d/2}} \prod_{i=1}^{n} \frac{\Gamma(\d[i])}{((\ell-p_i)^2)^{\d[i]}} 
\end{equation}
where  $\mathcal{A}$ is a numerical factor proportional to the coupling constant that we shall drop henceforth.

We Schwinger parametrize \eqref{1loop}, perform the resulting Gaussian integral over the loop momentum $\ell$, and find 
\begin{equation}\label{1lsch}
	\begin{split}
		{\cal M}_{\d[1],\ldots,\d[n]}(p_1,\ldots,p_n) 
		&= \prod_{i=1}^n \int_{0}^{\infty} dt_i {t_i^{\Delta_i-1} \over T^{d/2}} \exp(-{1\over T} \sum\limits_{\substack{i,j=1 \\ i \le j}}^{n} t_i t_j P_{ij}).
	\end{split}
\end{equation}
with the definitions $P_{ij}=  (p_i-p_j)^2$ and $T=\sum_{i=1}^n t_i$.

If each $P_{ij}$ is replaced by the corresponding position space invariant $s_{ij}=(x_i-x_j)^2$, then\eqref{1lsch} is seen to be equivalent to the contact contribution of the $n$-point massless correlator in position space. The equivalence enables us to translate the results of \cite{Prabhu:2023oxv} concerning massless field correlators in position space directly into momentum space. In particular, in \cite{Prabhu:2023oxv}, we showed that the $n$-point contact correlator of massless fields can be expressed as a conformal integral with $(n+1)$-points in the limit that the $(n+1)^{th}$ point is taken to infinity. This translates to the result that in momentum space, any $1$-loop diagram with $n$ internal massless fields can be written as a conformal integral i.e. one that corresponds to a $1$-loop diagram with $(n+1)$ internal lines whose scaling dimensions satisfy the conformal constraint $\sum_{i=1}^{n+1} \d[i]=d$. This result can be expressed as
\begin{equation}\label{1llim}
\begin{split}
\lim_{p_{n+1}^2 \rightarrow \infty} \amc{n+1}=\frac{ \Gamma(\d[n+1])}{(p_{n+1}^2)^{\d[n+1]}} \am{n},
\end{split}
\end{equation}
	where $\{\p_{ij}\}_{(n)}$ denotes the set of all momenta invariants $P_{ij}$ for $i,j=1, \ldots,n$.
	
The point at infinity can be brought to a finite point as seen in appendix \S \ref{sec:poscont} so that we find an equality valid for arbitrary kinematics, by translating \eqref{conteq} to momentum space to find
\begin{equation}\label{1leq}
\amp{n}{\frac{P_{ij}}{P_{i(n+1)}P_{j(n+1)}}}= \frac{\prod_{i=1}^{n} P_{i(n+1)}^{-\d[i]}}{\Gamma(\d[n+1])} \amc{n+1} 
\end{equation}
\begin{figure}
\centering  \includegraphics[width=.5\textheight]{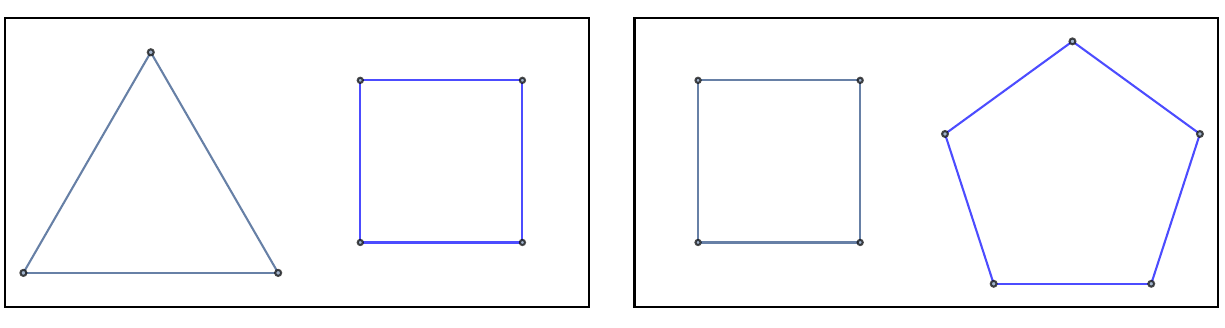}
\caption{Examples of 1 loop conformal families with the conformal diagrams in blue. }
\label{fig:1loop}
\end{figure}
Diagrammatically, this relates the 1-loop diagram with $n$ internal lines having general scaling dimensions to the conformal 1-loop diagram with $(n+1)$ internal lines. Following the arguments in \S \ref{sec:poscont}, it is clear that knowing the value of either of these diagrams fixes the other one completely, including prefactors. We will find answers to the diagrams of Fig. \ref{fig:1loop} in \S \ref{sec:conffam}.

 We now proceed to families with higher number of loops.
 \section{Families of L-loop diagrams}\label{sec:lloop}

 Consider the general $L-$loop integral in \eqref{feynint}. Every loop integral over some loop momentum $l_r$ can be expressed in the $1-$loop form in \eqref{1loop} with all the loop momenta other than $l_r$ entering into the definition of $p_i$. Introducing one extra internal line in each loop via \eqref{1llim} so that the conformal constraint is satisfied in the loop, one finds that the general $L-$loop Feynman diagram is given by an $L-$fold limit of a conformal integral. 
 
 Let us consider 2-loop diagrams first. We find the general 2-loop diagram to be related to the conformal diagram as
 
 \begin{equation} \label{2llim}
 	\ame{\ell}{\ell+1}{n}=\lim\limits_{p_{\ell+1}^2,p_{n+2}^2\to \infty}  {(p_{\ell+1}^2)^{\d[\ell+1]} (p_{n+2}^2)^{\d[n+2]}\over \Gamma(\d[\ell+1]) \Gamma(\d[n+2])} \amec{\ell+1}{\ell+2}{n+2} 
 \end{equation}
 where the general 2-loop diagram on the left is seen to be related to the conformal 2-loop diagram on the right hand side. The latter has a left loop with indices on the lines belonging to the set $L^{+}=\{0,1,\ldots,\ell+1\}$ and a right loop with indices belonging to $R^{+}=\{0,\ell+2,\ldots,n+2\}$. The sum of scaling dimensions of the lines in each loop, $\Delta_{L^{+}}$and $\Delta_{R^{+}}$ respectively, satisfy $\Delta_{L^{+}}=\Delta_{R^{+}}=d$, and so, this diagram is composed of two conformal integrals. Naturally, in general, there are two intermediate \emph{half conformal} diagrams composed of one conformal integral and one non-conformal integral. The one with the conformal integral on the left is related to the conformal 2-loop diagram by
 \begin{equation} \label{l2loop}
 	\ame{\ell+1}{\ell+2}{n+1}=\lim\limits_{p_{n+2}^2\to \infty}  {(p_{n+2}^2)^{\d[n+2]}\over  \Gamma(\d[n+2])} \amec{\ell+1}{\ell+2}{n+2} 
 \end{equation}
 As we did for the contact diagram in \S \ref{sec:poscont}, we move to position space, perform an inversion and translation by $x_{n+2}$, and convert back into momentum space to see 
 \begin{equation}\label{2leq}
 	\ampe{\ell+1}{\ell+2}{n+1}{\frac{P_{ij}}{P_{i(n+2)}P_{j(n+2)}}}= \frac{\prod_{i=1}^{n+1} P_{i(n+2)}^{-\d[i]}}{\Gamma(\d[n+2])} \amec{\ell+1}{\ell+2}{n+2}  
 \end{equation}
 
\begin{figure}
 	\includegraphics[width=.5\textheight]{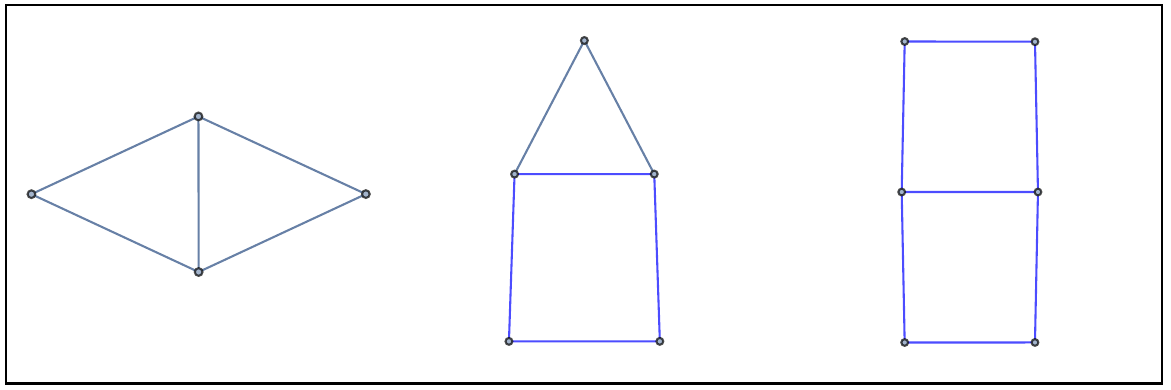}
 	\begin{center}
 		\caption{The conformal family of the conformal 2-box diagram with all blue lines. The middle diagram is, upto a prefactor, the same as the conformal 2-box on the right via \eqref{2leq} (with $\ell=3$ and $n=6$). The 2-triangle diagram on the left can be obtained from the conformal 2-box diagram via \eqref{2llim} (with $\ell=3$ and $n=6$). }
 		\label{fig:2tri}
 	\end{center}  
 \end{figure}

 Similarly, we can also relate the other half conformal diagram, the one with a conformal integral on the right, to the 2-loop conformal diagram. 
 At the 2-loop level then, we have three kinds of diagrams or rather, topologies. The conformal 2-loop diagram is equivalent to the half-conformal diagram via \eqref{2leq}, and can be used to obtain the general 2-loop diagram via \eqref{2llim}. Together , the three constitute a \emph{conformal family}. As an example, the family of the conformal 2-box diagram is depicted in Fig.\ref{fig:2tri}. There is one such conformal family for every possible conformal diagram, i.e. for every diagram composed of two polygons, each of which satisfy the conformal constraint.

 \begin{figure}
 	\centering  \includegraphics[width=.7\textheight]{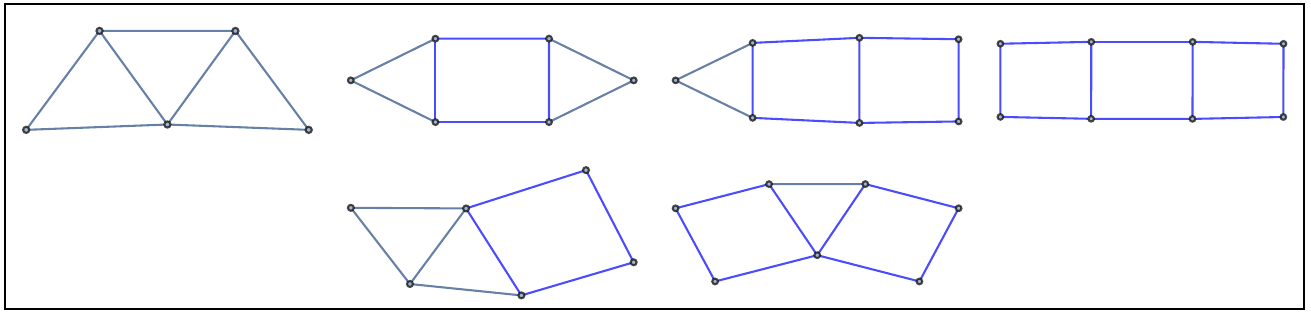}
 	\caption{The conformal family of the conformal 3-box diagram depicted with all blue lines.  The diagrams on the two right most columns are related to each other upto prefactors, and their appropriate limits provide the other diagrams. }
 	\label{fig:3box}
 \end{figure}

 \begin{figure}
 	\centering  \includegraphics[width=.6\textheight]{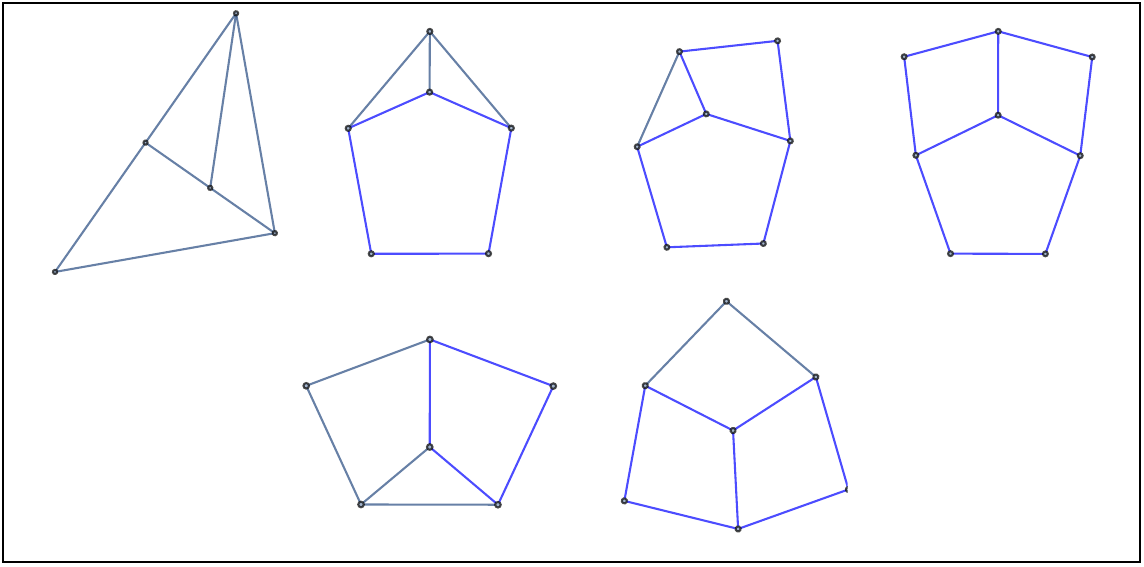}
 	\caption{The conformal family of the \emph{tennis court} diagram depicted with all blue lines.  }
 	\label{fig:tenc}
 \end{figure}

 The picture at two loops gets naturally generalized to the $L$- loop case. At each loop, there are several conformal families, one for each possible conformal diagram. Computing this conformal member in each family suffices to determine every other diagram in the family in the following way. Those members of the family that are related by contracting a single line of the conformal member are equivalent to the conformal member by a relation similar to \eqref{2leq}. Contracting more lines of the conformal member provides all other members of this family. We note that momentum conservation can be imposed on all diagrams thus obtained only if there is at least one external momentum flowing into each loop that is being contracted. In particular, this means that the total number of contractions possible is upper bounded by the number of external momenta. 
 
\begin{figure}
 	\centering  \includegraphics[width=.6\textheight]{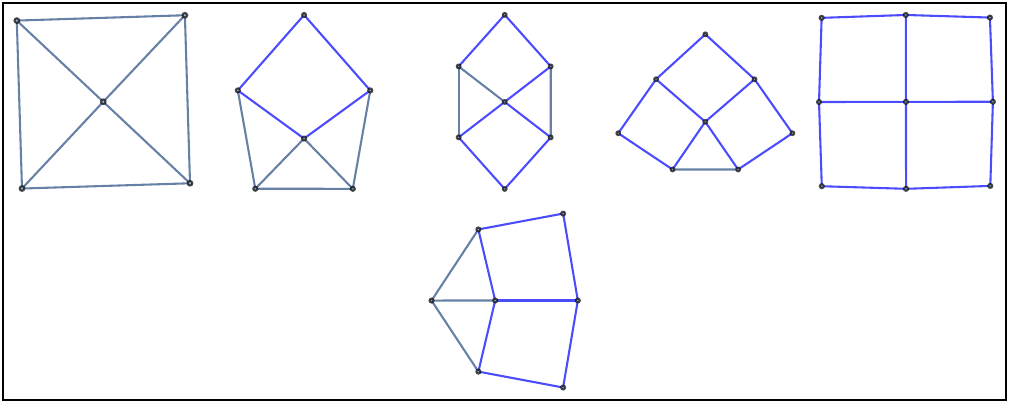}
 	\caption{The conformal family of the four loop \emph{net} diagram}
 	\label{fig:net}
 \end{figure}

 Some examples of families at three loops are listed in Figs. \ref{fig:3box} and \ref{fig:tenc}. We present an example of a family at four loops in Fig. \ref{fig:net}. In these figures, we depict all possible members of the family. Families with conformal members having lesser number of external fields than number of loops, will have lesser number of members.
 
\section{Ladder diagrams in four dimensions}\label{sec:lad}
Our relations between conformal and non-conformal diagrams includes, as a special case, ladder diagrams. In particular, consider massless ladder diagrams with an arbitrary number of rungs. The first of such diagrams are simply the non-conformal triangle and conformal box of Fig. \ref{fig:1loop} related by \eqref{1leq}. Ladder diagrams with two rungs are given by the two rightmost ones in Fig. \ref{fig:2tri}, and the relation between them in \eqref{2leq}. 

Let us consider a specific subclass of such ladder diagrams. With one external field each at the front and rear ends of the ladders, and unit propagator powers, these diagrams can be thought of as the 1-loop and 2-loop contributions to the 3-point and 4-point function respectively in $\phi^3$ theory in four dimensions. For this special case\footnote{They also showed the relation for some generalizations of the propagator powers/scaling dimensions that sum up to four.}, Davydychev and Usyukina \cite{Usyukina:1992jd} showed that the equivalences in \eqref{1leq} and \eqref{2leq} hold. They also generalized these relations to arbitrary loop order and found exact answers for them in \cite{Usyukina:1993ch}. 

The $4-$point $L-$loop conformal ladder diagram with all boxes \cite{Usyukina:1993ch} can be represented as 
\begin{equation}
	\mathcal{M}_{L-ladder}^{4-point}(z,\zbar)=\frac{1}{P_{23} P_{14}} \frac{1}{(z-\zbar)}	\sum_{m=L}^{2L} \frac{(-1)^m m! \log^{2L-m}{(z \zbar)}}{L!(m-L)! (2L-m)!} \left(\text {Li}_m(z)-\text{Li}_m(\zbar)\right)
\end{equation}
where $z,\zbar$ are related to the conformal cross ratios  \begin{equation} \label{cr} 
	u=\frac{P_{12}P_{34}}{P_{23}P_{14}}, \qquad  \qquad v=\frac{P_{24}P_{13}}{P_{23}P_{14}}
\end{equation}
via the relations
\begin{equation}
	u=z \zbar, \qquad v=(1-z)(1-\zbar)
\end{equation}
We use these exact answers to predict the answers to all the diagrams in the conformal families of the ladder diagrams. For $L\geq2$, given that these have two external fields in each of the loops at the ends, there are two possible contractions (without getting bubbles), and so two members. The conformal member above is equivalent to the $L-$loop contribution to the 3-point function given by the diagram with one triangle and $L-1$ boxes given by
\begin{equation} \label{3ptllad}
	\mathcal{M}_{L-ladder}^{3-point}(w,\bar w)=\frac{1}{P_{23}}\frac{1}{(w-\bar{w})}	\sum_{m=L}^{2L} \frac{(-1)^m m! \log^{2L-m}{(w \bar{w})}}{L!(m-L)! (2L-m)!} \left(\text {Li}_{m}(w)-\text{Li}_{m}(\bar{w})\right)
\end{equation}
where $w,\bar{w}$ are related to the simple ratios formed out of distances between the three external points 
\begin{equation}\label{sr3}
	r_1=\frac{P_{12}}{P_{23}}, \quad r_2=\frac{P_{13}}{P_{23}}	
\end{equation}
via the relations
\begin{equation}
	r_1=w \bar{w}, \qquad r_2=(1-w)(1-\bar{w})
\end{equation}
 The $3-$point function in \eqref{3ptllad} matches with the direct computation in \cite{Usyukina:1993ch}. Finally, the third member of the family is the $L$-loop contribution to a 2-point function with two triangles and $L-2$ boxes. This is given by the limit of $w, \bar{w} \rightarrow 1$ in the above expression. The combinations of  polylogarithms in \eqref{3ptllad} are all single valued at this point. Hence, we find that the arbitrary loop contribution to the massless two point function coming from the ladder diagrams is finite and is given by 
 \begin{equation} \label{2ptfin}
 	\mathcal{M}_{L-ladder}^{2-point}= \frac{(2L)!}{(L!)^2} \zeta(2L-1)
 \end{equation}
 The two loop contribution $6 \zeta(3)$ is well known\footnote{This value was derived in \cite{Chetyrkin:1980pr}. See \cite{Bierenbaum:2005zz} for a review of the 2-loop 2 point function.} The three and four loop results in \eqref{2ptfin} match with the values in \cite{Baikov:2010hf}.

\section{Conformal families in any dimension}\label{sec:conffam}
The non-conformal members of any conformal family can be obtained in general dimensions as an appropriate limit of the conformal member.  In momentum space, the non-conformal members are obtained by collapsing the edges of the conformal member which identifies a number of vertices. The momentum flowing into the new vertex is the sum of the momenta flowing into the vertices before the collapse. This is the translation into momentum space of the position space limit, obtained by taking one extra field at each vertex in the corresponding conformal diagram to infinity. This procedure gives us the answers to Feynman integrals in general dimensions and with general values of the propagator powers. In particular, for divergent diagrams, we can expand in the dimensional regularization parameter  around any dimensions for any values of the propagator powers.

\subsection{1-loop diagrams}
\begin{figure}
	\centering  \includegraphics[width=.3\textheight]{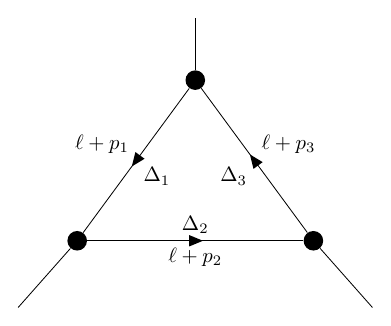}
	\caption{The 3-point triangle diagram with loop momentum $\ell$, momenta of internal lines as labelled and powers of corresponding propagators denoted by $\d[i]$}
	\label{fig:tri}
\end{figure}
The first non-trivial case is the equality between the triangle diagram with general scaling dimensions, and the conformal box diagram (i.e. the box diagram with scaling dimensions that satisfy the conformal constraint). The conformal box diagram in momentum space is the same as the conformal 4-point contact diagram in position space. As we reviewed in  \cite{Prabhu:2023oxv}, an exact answer to the massless four point conformal contact diagram can be found in position space. This is calculated using the methods of \cite{Symanzik:1972wj} to treat conformal integrals. The same integral also appears in several other places, such as the calculation of 4-point conformal blocks \cite{Dolan:2000ut}.  Translated into momentum space, this gives the answer to the conformal four point box integral. 

Using \eqref{1leq}, we see that it provides  the general triangle Feynman integral with arbitrary propagator powers in Fig. \ref{fig:tri} to be
\begin{equation} \label{tri}
	\begin{split}	 
	\mathcal{M}_{\d[1],\d[2],\d[3]}(p_1,p_2,p_3)=	\frac{\p_{23}^{\Db[3]}}{\Gamma(\Dt[3])} & \Bigg\{\Gamma(\Delta_1) \Gamma(-\Db[3]) \Gamma(\db[13]) \Gamma(\db[12]) \\
		& \times F_4(\Delta_1,-\Db[3],1-d/2-\Delta_{13},1-d/2-\Delta_{12};r_1,r_2) \\
		+&r_1^{\db[13]} r_2^{\db[12]}\, \Gamma(\Dt[3])\Gamma(-\db[1])\Gamma(-\db[12])\Gamma(-\db[13]) \\
		& \times F_4(\Dt[3],-\db[1],1+d/2+\Delta_{13},1+d/2+\Delta_{12};r_1,r_2)\\
		+& \Big( r_2^{\db[12]} \Gamma(\Delta_3) \Gamma(-\db[2])  \Gamma(\db[13])\Gamma(-\db[12]) \\
		&\times  F_4(\Delta_3,\db[2],1-d/2-\Delta_{13},1+d/2+\Delta_{12};r_1,r_2) +(2 \leftrightarrow 3) \Big) \Bigg\},\\ 
	\end{split}
\end{equation}
as a function of the simple ratios in \eqref{sr3}. This is the exact answer for the off-shell triangle diagram with all internal fields being massless and the external ones being massive or massless. As an example, the answer to the on-shell triangle diagram with three different external massive fields can be obtained by having $\p_{12}=m_1^2,\p_{23}=m_2^2$, and $\p_{31}=m_3^2$.
This means that we can use it to find the triangle diagram in dimensional regularization to arbitrary order in the regularization parameter. 

\begin{figure}
	\centering  \includegraphics[width=.3\textheight]{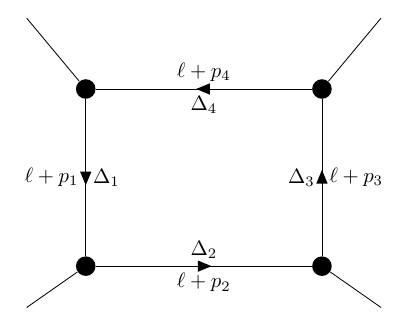}
	\caption{The 4-point box diagram}
	\label{fig:box}
\end{figure}

The next example is the relation between the general box diagram and the conformal pentagon diagram, depicted in Fig. \ref{fig:1loop}. The 4-point box diagram, shown in Fig. \ref{fig:box} is similar to the 4-point position space contact diagram, and so is given by a five-fold Mellin Barnes integral, as shown for example in \cite{Prabhu:2023oxv}.
This can be used to find the dimensionally regularized version of the box integral in any number of dimensions. For example, in 4 dimensions, we put $\Delta_i=1$ and $d=4-2\epsilon$ in \eqref{dcft5} and use \eqref{1leq} to find that the off-shell box integral is given by
\begin{equation} \label{box4dirdm}
	\begin{split}
	\mathcal{M}_{1,1,1,1;4-2\epsilon}(p_1, p_2, p_3,p_4)= {1 \over \Gamma(-2\epsilon) }  {1 \over \p_{24}^{\epsilon-1}} \,& \prod_{i=1}^{5}\int  {da_i \over {2\pi i}}   \,r_i^{a_i}  \,\Gamma(-a_i)  \Gamma(1+a_{245})\Gamma(1+a_{123})   \\
		&\times \Gamma(-1-\epsilon-a_{124}) \Gamma(-1-\epsilon-a_{235} )\Gamma(2+\epsilon+\A[5]) \\
	\end{split}
\end{equation}
where we have defined the symbols $a_{i_1,i_2,\ldots,i_n}=a_{i_1}+a_{i_2}+\cdots a_{i_n},\A[n]=\sum_{i=1}^{n}a_i$ and the ratios $r_i$ are given by
\begin{equation}
	r_1=\frac{\p_{14}}{\p_{24}}, r_2=\frac{\p_{13}}{\p_{24}}, r_3=\frac{\p_{12}}{\p_{24}},r_4=\frac{\p_{34}}{\p_{24}},r_5=\frac{\p_{23}}{\p_{24}}
\end{equation}
The exact result is in terms of a five-fold Mellin integral and can be expressed in terms of five variable hypergeometric functions. Moreover, the expression in \eqref{box4dirdm} can be expanded to arbitrary order in the dimensional regularization parameter $\epsilon$.

We now look at the on-shell box diagram. Imposing momentum conservation, the external momenta are differences of the kind $p_i-p_j$, which we can choose to be cyclic. In the limit that all the external momenta are on-shell, we find that only one of the five ratios is non-zero so that we have
\begin{equation} \label{boxgendim}
	\begin{split}
		& \mathcal{M}_{\d[1],\ldots,\d[4]}^{on-shell}(r)	=\frac{ \p_{24}^{\Db[4]}}{\Gamma\left(\Dt[4]\right)} \mi[a] \, r^{a} \Gamma(-a) \Gamma(\d[3]+a)\Gamma(\d[4]+a)\Gamma\left(-\Db[4]+a\right)\Gamma\left(\db[134]-a\right)\Gamma\left(\db[234]-a\right) \\
		=&\frac{ \p_{24}^{\Db[4]}}{\Gamma\left(\Dt[4]\right)}  \Bigg\{\Gamma(\d[3])\Gamma(\d[4])  \Gamma\left(-\Db[4]\right) \Gamma\left(\db[134]\right)\Gamma\left(\db[234]\right)\  \pFq{3}{2}{\d[3],\d[4],-\Db[4]}{-\db[134]+1,-\db[234]+1}{-r} 		\\
		+&\left(\Gamma(\d[1])\Gamma(\db[23])\Gamma(\db[24])  \Gamma\left(\d[2,1]\right) \Gamma\left(-\db[234]\right) r^{\db[234]} \, \pFq{3}{2}{\d[1],\db[23],\db[24]}{1-\d[2,1],1+\db[234]}{-r}		+(1\leftrightarrow 2)\right) \Bigg\},
	\end{split}
\end{equation}
where the ratio $r=\frac{\p_{13}}{\p_{24}}$. This expression matches with Eq. 3.18 of \cite{Tarasov:2017jen} where the on-shell box with arbitrary propagator powers was found via the method of differential equations. In particular, we can use this expression to find the dimensionally regularized on-shell box in any dimensions, for example the $d=4-2\epsilon$ result in \cite{Bern:1993kr}.

\subsection{2-loops}
\begin{figure}
	\centering  \includegraphics[width=.4\textheight]{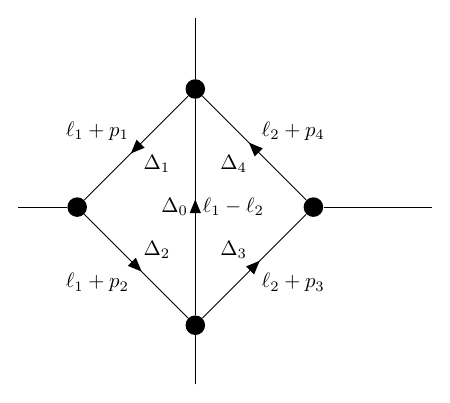}
	\caption{The $4-$point \emph{kite} diagram, with the momenta in the two loops being $\ell_1$ and $\ell_2$.}
	\label{fig:kite}
\end{figure}
We move on to the $2-$loop family of the conformal double box diagram depicted in Fig. \ref{fig:2tri}. We want to compute the 4-point double triangle box diagram i.e the leftmost diagram in Fig. \ref{fig:2tri}. In what follows, for brevity, we shall refer to this diagram as the 4-point kite diagram depicted once again in Fig. \ref{fig:kite}. We calculate it by implementing the limit in \eqref{2llim} in position space as follows. The 6-point conformal double box in momentum space is the same as the 6-point conformal exchange diagram in position space treated, for example, in \cite{Prabhu:2023oxv}. We first take the limit where one external field on the left vertex, is coincident with another on the right vertex before taking this coincident position to infinity. The coincident limit ($6\rightarrow 5$) in position space takes us to the 5-point conformal double box integral in momentum space given by the seven-fold Mellin integral 
\begin{equation}\label{ckite5}
	\begin{split}
		\mathcal{M}_{\d[1],\ldots,\d[6]}^{cft}=&\Gamma(\d[5])	 \left(\frac{\p_{35} \p_{15}}{\p_{13}}\right)^{\dt[506]} \frac{1}{\p_{15}^{\d[1]} \p_{25}^{\d[2]} \p_{35}^{\d[3]}\p_{45}^{\d[4]}} \\
		&\times \prod_{i=1}^{5} \mi[a_{i}]  \Gamma(-a_i) s_i^{a_i}\mi[p]  \mi[q] \, \frac{\Gamma(-p)\Gamma(-q)}{\prod_{i=3}^5 \Gamma(\d[i]')} s_5^q \, \mdcft_{\d[0],\d[3],\d[4],\d[6]}(p,q) \mdcft_{\d[1],\d[2],\d[3]',\d[4]',\d[5]'}(a_1,\ldots,a_5) 
	\end{split}
\end{equation}
with the definitions
\begin{equation}
	\d[i]'= \begin{cases}
		-p & i=3 \\
		p+q+\db[6] & i=4 \\
		-q-\db[506] & i=5
	\end{cases}
\end{equation}
and the choice of the five cross ratios of five points,
\begin{equation} \label{cr5}
	s_1=\frac{\p_{12}\p_{35}}{\p_{13}\p_{25}}, s_2=\frac{\p_{23}\p_{15}}{\p_{13}\p_{25}},s_3=\frac{\p_{15}\p_{35}\p_{24}}{\p_{13}\p_{25}\p_{45}},s_4=\frac{\p_{14}\p_{35}}{\p_{13}\p_{45}},s_5=\frac{\p_{34}\p_{15}}{\p_{13}\p_{45}}
\end{equation}
Now, implementing the second step of taking the coincident position to infinity lands us at the fully off-shell $4-$point kite diagram in Fig. \ref{fig:kite}. This involves replacing the conformal cross ratios $s_i$ in \eqref{ckite5} with the corresponding simple ratios $r_i$ given by
\begin{equation} \label{sr5}
	r_1=\frac{\p_{12}}{\p_{13}}, r_2=\frac{\p_{23}}{\p_{13}},r_3=\frac{\p_{24}}{\p_{13}},r_4=\frac{\p_{14}}{\p_{13}},r_5=\frac{\p_{34}}{\p_{13}}
\end{equation}
and making use of the conformal constraints 
\begin{equation}
	\d[0346]=\d[0125]=d
\end{equation}
to find
\begin{equation}\label{4kiteoff}
	\begin{split}
		\mathcal{M}_{off-shell}^{kite}=&\Gamma(\dt[012])	\, \p_{13}^{\dt[01234]}  \, \prod_{i=1}^{5} \mi[a_{i}]  \Gamma(-a_i) s_i^{a_i}\\
		&\times\mi[p]  \mi[q] \, \frac{\Gamma(-p)\Gamma(-q)}{\prod_{i=3}^5 \Gamma(\d[i]')} s_5^q \, \mdcft_{\d[0],\d[3],\d[4],\d[6]}(p,q) \mdcft_{\d[1],\d[2],\d[3]',\d[4]',\d[5]'}(a_1,\ldots,a_5) 
	\end{split}
\end{equation}
where 
\begin{equation}
	\d[i]'= \begin{cases}
		-p & i=3 \\
		p+q-\db[034] & i=4 \\
		-q+3d/2-\d[01234] & i=5
	\end{cases}
\end{equation}
 The Mellin integral over $p$ can be done using Barnes first lemma. Performing the Mellin integral over $q$ then gives us a five-fold Mellin Barnes integral representation whose coefficients involve a hypergeometric series in the cross ratio $s_5$. This can be used to find a five-fold series representation for this integral.

Here, instead, we shall find the on-shell version of the $4-$point kite diagram by choosing the cyclic differences in momenta to be zero. This collapses the $q$ integral and four of the $a_i$ integrals. Performing the $p$ integral lands us at a single Mellin Barnes integral given by

\begin{equation}\label{4kiteon}
	\begin{split}
		& 	\frac{\Gamma(\db[12]) \Gamma(\db[34]) \Gamma(\db[0])}{\Gamma(\dt[012]) \Gamma(3d/2-\d[01234])} \frac{1}{\p_{13}^{\dt[01234]}} \\
		& \times \mi[a_3] \Gamma(-a_3) r_3^{a_3} \Gamma(\d[2]+a_3)\Gamma(\d[4]+a_3)\Gamma(\dt[01234]+a_3)\Gamma(\dt[0234]-a_3)\Gamma(\dt[0124]-a_3)\\
		=&\frac{\Gamma(\db[12]) \Gamma(\db[34]) \Gamma(\db[0])}{\Gamma(\dt[012]) \Gamma(3d/2-\d[01234])} \frac{1}{\p_{13}^{\dt[01234]}} \\
		& \times \Bigg\{\Gamma(\d[2])\Gamma(\d[4])  \Gamma\left(\dt[01234]\right) \Gamma\left(\dt[0234]\right)\Gamma\left(\dt[0124]\right)\  \pFq{3}{2}{\d[2],\d[4],\dt[01234]}{-\dt[0234]+1,-\dt[0124]+1}{-r_3} 		\\
		+&\left(\Gamma(\d[1]+2\d[0234])\Gamma(-\dt[0234])\Gamma(\dt[023])  \Gamma\left(\d[3,1]\right) \Gamma\left(\dt[034]\right) r^{\dt[0234]} \, \pFq{3}{2}{\dt[023],\dt[034],\d[1]+2\dt[0234]}{1-\Delta_{3,1} ,1+\dt[0234]}{-r_3}		+(1\leftrightarrow 3)\right) \Bigg\},
	\end{split}
\end{equation}
The expressions in \eqref{4kiteoff} and \eqref{4kiteon} can be used to find the dimensionally regularized version of the off-shell and on-shell versions respectively of the $4-$point kite integral for any arbitrary powers of the propagators and in any dimension to all orders in the dimensionally regularized parameter.

\section{Conclusions and Discussion}\label{sec:conc}
In this article, we considered general multi-loop Feynman integrals in arbitrary dimensions with massless internal fields and massive or massless external ones. However, the idea that they can be derived from conformal integrals applies to Feynman diagrams and integrals coming from massive internal lines as well. The corresponding diagrams, are of course, more complicated compared to the massless ones. We will report on such relations between Feynman integrals with massive internal fields elsewhere. 

Conformal integrals have been the subject of study from many different points of view \footnote{While they are usually studied in position space, they have also been more recently computed in momentum space in \cite{Bzowski:2013sza,Bzowski:2015yxv,Bzowski:2020kfw}.}. After the recent revival of the conformal bootstrap \cite{Rattazzi:2008pe,Poland:2018epd}, the computation of conformal blocks and conformal integrals has been vigorously pursued \cite{Costa:2011dw,Simmons-Duffin:2012juh,Hogervorst:2013sma,Alkalaev:2015fbw,Karateev:2017jgd,Rosenhaus:2018zqn, Parikh:2019ygo,Fortin:2019dnq,Parikh:2019dvm,Fortin:2019zkm,Pal:2020dqf,Buric:2020dyz,Buric:2021ywo,Pal:2021llg,Pal:2023kgu}. 
It would be interesting and useful to pursue this connection to help with the computation of multi-loop Feynman diagrams. The knowledge of the full conformal integral gives us the off-shell version of the Feynman diagrams, 
whereas computing it in the limit where some cross ratios are zero suffices to compute on-shell Feynman diagrams. We used this idea to compute the on-shell versions of two 4-point diagrams, the 1-loop box and the 2-loop kite. It would be interesting to take this program further, and compute exact answers to the other conformal families, the simplest of which are the ladder diagram families. 

We used the Mellin Barnes integral representation to actually compute the conformal integrals. For such an $n-$point contact conformal diagram, the number of Mellin integrals needed to compute it is the number of cross ratios that can be formed from $n$ points. This bounds from below the number of Mellin integrals needed for any $n-$point diagram. The latter increases as the diagrams get more complicated, and seems to depend on the choice of cross ratios. It would be great to find a procedure that results in the least number of Mellin integrals for every diagram. Very recently, tremendous progress has been made in the computation of multi-fold Mellin integrals in \cite{Ananthanarayan:2020fhl,Ananthanarayan:2020ncn,Banik:2022bmk,Banik:2023rrz,Banik:2024ann}, and they also provide packages and Mathematica implementations to convert these integrals into series solutions. It would be of interest to use these to provide explicit answers for the non-conformal diagrams of each family. For example, in \cite{Ananthanarayan:2020ncn} the explicit series solutions in general dimensions and with arbitrary propagator powers have been found for two conformal diagrams at 6-points, namely the double box and the hexagon diagrams. These readily provide series solutions for all the Feynman diagrams in the respective families explicitly, which can be used to find their dimensionally regularized versions.

 Feynman integrals have received an immense amount of attention both for their computational importance as well as for their mathematical structure. It might be useful to adapt some of these methods to the study of conformal integrals. In particular, it would also be interesting to see if we can develop the ideas presented here to help with the organization and computation of conformal blocks by showing its equivalence to a set of Feynman integrals derived from the usual Feynman rules. In this context, we note that a diagrammatic representation and Feynman-like rules for computing conformal blocks have been presented in \cite{Hoback:2020pgj,Fortin:2020ncr,Fortin:2022grf}. 

Here, we have used conformal integrals to find Feynman integrals exactly, and so also in dimensional regularization.  We could also ask if there is a way to regularize the conformal integrals themselves. One might worry that a regularization scheme such as dimensional regularization will spoil the conformal invariance properties we have described. However, it has been shown, in the case of dual conformally invariant integrals, that a regularization scheme exists which preserves dual conformal invariance\cite{Bourjaily:2013mma,Bourjaily:2019jrk,Bourjaily:2019vby}. Another possibility is to use a modified version of dimensional regularization where the conformal constraints continue to be satisfied by suitable modifications to the scaling dimensions appearing in the integrals too. We will pursue this approach elsewhere.

\section*{Acknowledgments}
It is a pleasure to thank Matt von Hippel, Alok Laddha, Gautam Mandal, Shiraz Minwalla, Sarthak Parikh and Onkar Parrikar for valuable discussions. We would also like to thank all the participants of a virtual tea discussion at TIFR where some parts of this work were presented. A preliminary version of this work was also presented at the conference "Future Perspectives on QFT and Strings" at IISER Pune. We would like to thank the organizers of this conference for their hospitality, and the audience for many discussions. This work was supported by SERB, Department of Science and Technology under the grant PDF/2021/004777. We also acknowledge support from a J. C. Bose Fellowship JCB/2019/000052, and from the Infosys Endowment for the study of the Quantum Structure of Spacetime. Research at TIFR is supported by the Department of Atomic
Energy under Project Identification No. RTI4002. We would like to express our gratitude to the people of India for their support in the development of the basic sciences.

	\appendix
	\section{Notation}
	We lay out some convenient notation used in this article. For functions of scaling dimensions $\d[i]$, we use the following definitions
	\begin{equation}
		\begin{split}
			\d[ij]&=\d[i]+\d[j] \\
			 \d[ij,k]&=\d[i]+\d[j]-\d[k] \quad \text{etc.}\\
			 \db[i]&=d/2-\d[i] \\
			 \dt[i]&=d-\d[i] 	\\
			\D[n]&=\sum_{i=1}^{n} \d[i]	
		\end{split}
	\end{equation}
	
	We define the standard measure for Mellin integrals to be 
	\begin{equation}
		\mi[a]=\int_{-i \infty}^{+i \infty} \frac{da}{2\pi i}
	\end{equation}

	\section{Position space contact diagrams}\label{sec:poscont}
		As discussed in \cite{Prabhu:2023oxv}, the contact diagram for $n$ massless fields can be expressed as
	\begin{equation}
		\gflata[1,n][1,n]=	\mathcal{G}^{\flat}_{\d[1],\ldots,\d[n]}(\{s_{ij}\}_{(n)})=H_{\d[1],\ldots,\d[n]}(\{s_{ij}\}_{(n)}) \mathcal{D}^{\flat}_{\d[1],\ldots,\d[n]}(\{r_{i}\}_{(n)})
	\end{equation}
	where $\{s_{ij}\}_{(m)}$ denotes the set of all distance invariants $s_{ij}=(x_i-x_j)^2$ for $i,j=1, \ldots,m$. The $H$ functions are simple products of powers of $s_{ij}$ that carry the scale, and the $\dflat$ functions, introduced in \cite{Pibv}, depend on the scaleless $r_n$ number of simple ratios of distances, denoted by $r_i$, that can be formed from $n$ points. 
	The $\dflat$ functions can be represented as $r_n$ fold Mellin integrals in the form
	\begin{equation}
		\mathcal{D}^{\flat}_{\d[1],\ldots,\d[n]}(\{r_{i}\}_{(n)})=	\prod_{i=1}^{r_n} \mi[a_i] \, r_i^{a_i} \, \Gamma(-a_i) \tilde{\mathcal{D}}_{\d[1],\ldots,\d[n]} (a_{1},\ldots,a_{r_n})
	\end{equation}
	In \cite{Prabhu:2023oxv}, it was shown that the contact diagram for $n$ massless fields can be expressed as a conformally invariant contact diagram for $(n+1)$ fields $\mathcal{G}^{cft}_{\d[1],\ldots,\d[n+1]}$ as
	\begin{equation}\label{poscont}
		\mathcal{G}^{\flat}_{\d[1],\ldots,\d[n]}(\{s_{ij}\}_{(n)})=\lim_{x_{n+1}^2\rightarrow\infty} \frac{(x_{n+1}^2)^{\d[n+1]}}{\Gamma(\d[n+1])} \ 	\mathcal{G}^{cft}_{\d[1],\ldots,\d[n+1]}(\{s_{ij}\}_{(n+1)})
	\end{equation}
	
	The conformally invariant $\gcft$ function takes the form 
\begin{equation}
		\gcfta[1,n][1,n]=	\mathcal{G}^{cft}_{\d[1],\ldots,\d[n]}(\{s_{ij}\}_{(n)})=H^{cft}_{\d[1],\ldots,\d[n]}(\{s_{ij}\}_{(n)}) \mathcal{D}^{cft}_{\d[1],\ldots,\d[n]}(\{s_{i}\}_{(n)})
\end{equation}
    where now the $\dcft$ functions depend on the $c_n$ number of cross ratios of distances that can be formed from $n$ points, represented as $c_n$ fold Mellin integrals in the form
	\begin{equation}
	\mathcal{D}^{cft}_{\d[1],\ldots,\d[n]}(\{r_{i}\}_{(n)})=	\prod_{i=1}^{c_n} \mi[a_i] \, s_i^{a_i} \, \Gamma(-a_i) \tilde{\mathcal{D}}^{cft}_{\d[1],\ldots,\d[n]} (a_{1},\ldots,a_{c_n})
	\end{equation}
    In \cite{Prabhu:2023oxv}, it was observed that the number of simple ratios that can be formed from $n$ points $r_n$ is the same as $c_{n+1}$, the number of cross ratios that can be constructed from $(n+1)$ points, 
    \begin{equation}
    	r_n=c_{n+1}= \begin{cases}
    		\frac{(n+1)(n-2)}{2} &  \quad n\leq d+1 \\
    		(n+1)d-\frac{(d+1)(d+2)}{2} & \quad  n >d+1   		
    	\end{cases}
    \end{equation}
	Let us write down the $\gcft$ functions at four and five points, derived in \cite{Prabhu:2023oxv}, in the form used in the main text. At four points, we choose the two cross ratios to be
	\begin{equation}
		s_1= \frac{s_{14}s_{23}}{s_{13}s_{24}}  \qquad  s_2= \frac{s_{12}s_{34}}{s_{13}s_{24}}
	\end{equation}
and can represent the conformally invariant function through
	\begin{equation}
	\begin{split}
			\hc{1}{4}&=\left(\frac{s_{14}s_{34}}{s_{13}}\right)^{\db[4]} \frac{1}{\prod_{i=1}^3 s_{i4}^{\d[i]}} \\
			\dc{1}{4}{2}&=\Gamma(-\db[14]-a_1) \Gamma(-\db[34]-a_2) \Gamma(\d[2]+a_{12}) \Gamma(\db[4]+a_{12})
	\end{split}
	\end{equation}
	At five points, with the choice of conformal cross ratios given by 
	\begin{equation} \label{cr5}
		s_1=\frac{s_{12}s_{35}}{s_{13}s_{25}}, s_2=\frac{s_{23}s_{15}}{s_{13}s_{25}},s_3=\frac{s_{15}s_{35}s_{24}}{s_{13}s_{25}s_{45}},s_4=\frac{s_{14}s_{35}}{s_{13}s_{45}},s_5=\frac{s_{34}s_{15}}{s_{13}s_{45}},
	\end{equation}
	we find
	\begin{equation} \label{dcft5}
		\begin{split}
			\hc{1}{5}&=\left(\frac{s_{15}s_{45}}{s_{14}}\right)^{\db[5]} \frac{1}{\prod_{i=1}^4 s_{i5}^{\d[i]}} \\
			\dc{1}{5}{5}&=\Gamma(-\db[15]-a_{235}) \Gamma(-\db[45]-a_{134})\Gamma(\d[2]+a_{123})\Gamma(\d[3]+a_{345})  \Gamma(\db[5]+a_{12345})
		\end{split}
	\end{equation}
	
	The conformal correlator in \eqref{poscont} has its $(n+1)^{th}$ point at infinity. Let us now bring this to a finite point by conformal transformations. In order to do this we first perform an inversion of both sides. The right hand side being a conformal correlator only picks up the product of conformal factors $(x^2)^{\Delta}$ for each field, whereas, on the left hand side, we transform the arguments using $s_{ij} \rightarrow \frac{s_{ij}}{x_i^2 x_j^2 }$. We then translate both sides by $x_{n+1}$ to find that
	\begin{equation}\label{conteq}
		\mathcal{G}^{\flat}_{\d[1],\ldots,\d[n]}\left(\left\{\frac{s_{ij}}{s_{i(n+1)}s_{j(n+1)}}\right\}_{(n)}\right)= \frac{\prod_{i=1}^{n} s_{i(n+1)}^{-\d[i]}}{\Gamma(\d[n+1])} \ 	\mathcal{G}^{cft}_{\d[1],\ldots,\d[n+1]}(\{s_{ij}\}_{(n+1)})
	\end{equation}
	In this form, it is entirely clear that the $n$-point contact correlator is upto, a simple prefactor, equal to the $(n+1)$-point conformal contact correlator for arbitrary kinematics. In \cite{Prabhu:2023oxv}, we showed that the the two correlators are equal in the so-called unit circumcentric configuration, from which we can also get to \eqref{conteq} by employing a conformal transformation to go to the general configuration. 
	
\bibliographystyle{JHEP.bst}
\bibliography{references.bib}
	
\end{document}